\documentclass[12pt]{article}
	\pdfoutput=1	
\usepackage{amsmath,amssymb}
\usepackage{graphicx}
\usepackage[pdftex]{hyperref}
\usepackage{caption}
\DeclareGraphicsExtensions{.eps,.bmp,.wmf,.jpg}
\numberwithin{equation}{section}
\def\be{\begin{equation}}
\def\ee{\end{equation}}

\def\bea{\begin{eqnarray}}
\def\eea{\end{eqnarray}}

\title{Dynamical analysis for a scalar-tensor model with Gauss-Bonnet and non-minimal couplings}
\author{L.N. Granda\thanks{luis.granda@correounivalle.edu.co} ,\, D. F. Jimenez\thanks{jimenez.diego@correounivalle.edu.co}\\{\it Departamento de Fisica, Universidad del Valle}\\{\it A.A. 25360, Cali, Colombia}}
\date{}
\begin{document}
\maketitle

\begin{abstract}
We study the autonomous  system for a scalar-tensor model of dark energy with Gauss-Bonnet and non-minimal couplings. The critical points describe important stable asymptotic scenarios including quintessence, phantom and de Sitter attractor solutions.  
Two functional forms for the coupling functions and the scalar potential were considered: power-law and exponential functions of the scalar field. For the exponential functions the existence of stable quintessence, phantom or de Sitter solutions, allows an asymptotic behavior where the effective Newtonian coupling becomes constant. The phantom solutions could be realized without appealing to ghost degrees of freedom. Transient inflationary and radiation dominated phases can also be described.
\begin{description}
\item[PACS numbers]
98.80.-k, 95.36.+x, 04.50.Kd
\end{description}
\end{abstract}

%\pacs{98.80.-k, 11.25.-w, 04.50.+h}% PACS, the Physics and Astronomy
\maketitle

%\tableofcontents

\section{\label{intro}Introduction}
The explanation of the late time accelerated expansion of the universe, confirmed by different observations  \cite{riess}, \cite{perlmutter}, \cite{kowalski}, \cite{hicken}, \cite{komatsu}, \cite{percival}, \cite{planck1}, \cite{planck2} represents one of the most important challenges of the modern cosmology. The current observational evidence for dark energy remains consistent with the simplest model of the cosmological constant, but there is no explanation to its smallness compared with the expected value as the vacuum energy in particle physics  \cite{starovinsky1}, \cite{peebles}, \cite{padmanabhan}. In addition, according to the analysis of the observational data, the equation of state parameter $w$ of the dark energy (DE) lies in a narrow region around  the phantom divide  ($w=-1$) and could even be below $-1$. All this motivates the study of alternative theoretical models, that give a dynamical nature to DE, ranging from a variety of scalar fields of different nature \cite{chiba}-\cite{sushkov1} to modifications of general relativity that introduce large length scale corrections explaining the late time behavior of the Universe \cite{capozziello1}-\cite{ tsujikawa} (see  \cite{copeland}-\cite{nojiriod} for review).\\
                                                            %%%%%%%%%%%%%%%%%%%%%%%
The low-energy limit of  fundamental physical theories like the string theory constitute an important source of physical models to address the dark energy problem. These string inspired models usually contain higher-curvature corrections to the scalar curvature term and direct couplings of the scalar fields to curvature \cite{metsaev}, \cite{meissner}. The couplings of scalar field to curvature also appear in the process of quantization on curved space time \cite{ford, birrel} and after compactification of higher dimensional gravity theories \cite{lamendola1}.
These couplings provide in principle a mechanism to evade the coincidence problem, allowing (in some cases) the crossing of the phantom barrier \cite{boisseau}, \cite{polarski}, \cite{perivolaropoulos}, \cite{fujii}. A representative model of this type of theories, subject of study in the present work, is the one that contains non minimal coupling to curvature and to the Gauss Bonnet (GB) invariant. The GB term is topologically invariant in four dimensions, but nevertheless it affects the cosmological dynamics when it is coupled to a dynamically evolving scalar field through arbitrary function of the field. In addition, this coupling has the well-known advantage of giving second order differential equations, preserving the theory ghost free. The role of the non-minimal coupling in the DE problem has been studied in different works, including the constraint on the coupling by solar system experiments
\cite{chiba}, the existence and stability of cosmological scaling solutions \cite{uzan, lamendola}, perturbative aspects and incidence on CMB \cite{perrotta, riazuelo}, tracker solutions \cite{perrotta1}, observational constraints and reconstruction \cite{boisseau, polarski, capozziello3} the coincidence problem \cite{tchiba}, super acceleration and phantom behavior \cite{vfaraoni}-\cite{polarski1}, asymptotic de Sitter attractors \cite{vfaraoni1} and a dynamical system analysis \cite{msami}.
%%%%%%%%%%%%%%%%%%%%%%%%%%%%%%%%%%%%%%%%%%%%%
On the other hand, the GB invariant coupled to scalar field has been  proposed to address the dark energy problem in \cite{sergei4}, where it was found that quintessence or phantom phase may occur in the late time universe. Different aspects of accelerating cosmologies with GB correction have been also discussed in \cite{tsujikawa1}, \cite{leith}, \cite{koivisto1}, \cite{koivisto2}, \cite{neupane2}, and a modified GB theory applied to dark energy have been suggested in \cite{nojiriod1}. For  a model with kinetic and GB couplings \cite{granda5}, solutions with Big Rip and Little Rip singularities have been found, and in \cite{granda6} the reconstruction of different cosmological scenarios, including known phenomenological models has been studied. In \cite{granda7} a model with non-minimal coupling to curvature and GB coupling was considered to study dark energy solutions, where a detailed reconstruction procedure was studied for any given cosmological scenario. In absence of potential  exact cosmological solutions were found, that give equations of state of dark energy consistent with current observational constraints.\\
Despite the lack of sufficient astrophysical data to opt for one or another model, it is interesting to consider scalar tensor couplings to study late time Universe since it could provide clues about how fundamental theories at high energies manifest at cosmological scales. The different studies of accelerating cosmologies with GB correction demonstrate that it is quite plausible that the scalar-tensor couplings predicted by fundamental theories may become important at current, low-curvature universe.\\
In the present paper we study the late time cosmological dynamics for the scalar-tensor model with non-minimal and Gauss-Bonnet couplings. To this end, and due to the non-linear character of the cosmological equations, we consider the autonomous system and analyze the cosmological implications derived from the different critical points.
The paper is organized as follows. In section II we introduce the model and give the general equations, which are then expanded on the FRW metric. In section III we introduce the dynamical variables, solve the equations for the critical points and give an analysis of the different critical points. In section IV we give a summary and discussion.
%%%%%%%%%%%%%%%%%%%%%%%%%%%%%%%%%%%%%%%%%%%%%%%%%%%%%%
\section{The action and field equations}
The action for the scalar field with non-minimal coupling of the scalar field to curvature and the coupling of the scalar field to the Gauss-Bonnet invariant, including also the matter content, is given by the equation (\ref{eq1}) below. The non-linear character of the cosmological equations  makes the integration of the same ones very difficult for a given set of initial conditions. Nevertheless the autonomous system for this model allows to study some interesting scaling solutions and the cosmological implications coming out from the different critical points.
\begin{eqnarray}\label{eq1}
S_{\phi}=&&\int d^{4}x\sqrt{-g}\Bigg[\frac{1}{2}F(\phi)R-\frac{1}{2}\partial_{\mu}\phi\partial^{\mu}\phi\nonumber\\
&&-V(\phi)-\eta(\phi){\cal{G}}+{\cal{L}}_m\Bigg],
\end{eqnarray}
where
\be\label{eq1a}
F(\phi)=\frac{1}{\kappa^2}-h(\phi),
\ee
${\cal{G}}$ is the Gauss-Bonnet invariant
\be\label{eq1b}
{\cal{G}}=R^2-4R_{\mu\nu}R^{\mu\nu}+R_{\mu\nu\lambda\rho}R^{\mu\nu\lambda\rho},
\ee
${\kappa}^2 =8\pi G$, ${\cal{L}}_m$ is the Lagrangian for perfect fluid with energy density ${\rho _m}$ and pressure $ {P_m}$, $h(\phi)$ and $\eta(\phi)$ are the non-minimal coupling and Gauss-Bonnet coupling functions respectively. Note that the coefficient of the scalar curvature $R$ can be associated with an effective Newtonian coupling as $\kappa_{eff}^2=F(\phi)^{-1}$. We will consider the spatially-flat Friedmann-Robertson-Walker (FRW) metric. 
\begin{equation}\label{eq2}
ds^2=-dt^2+a(t)^ 2\sum_{i=1}^{3}(dx_i)^2
\end{equation}
The cosmological equations with Hubble parameter $H=\dot{a}/a$ can be written in the form
%\begin{eqnarray}\label{eq3}
 % H^2 = &&\frac{\kappa^2}{3} \Bigg(\frac{1}{2} \dot{\phi}^2+V(\phi) + 6 \xi H \dot{\phi} \phi +
  %3\xi H^2 \phi^2\nonumber\\
%&&+24H^3\frac{d\eta}{d\phi}\dot{\phi}+{\rho_m }\Bigg)
%\end{eqnarray}
\be\label{eq3}
3H^2(F-8\dot{\eta}H)=\frac{1}{2}\dot{\phi}^2+V-3H\dot{F}+\rho_m
\ee
%\begin{eqnarray}\label{eq4}

\be\label{eq4}
2\dot{H}(F-8\dot{\eta}H)=-\dot{\phi}^2-\ddot{F}+H\dot{F}+8H^2\ddot{\eta}-8H^3\dot{\eta}
-(1+w_m)\rho_m
\ee
\be\label{eq5}
\ddot{\phi}+3H\dot{\phi}+\frac{dV}{d\phi}-3(2H^2+\dot{H})\frac{dF}{d\phi}+24H^2(H^2+\dot{H})\frac{d\eta}{d\phi}=0
\ee
\be\label{eq5a}
\dot{\rho_m}+3H\left(\rho_m+p_m\right).
\ee
This last equation is the equation for the perfect fluid that we use to model the matter Lagrangian.
Here the pressure $p_m=w_m\rho_m$, where $w_m$ is the constant equation of state (EoS) for the matter component.
The Eq. (\ref{eq3}) can be rewritten as
\be\label{eq6}
1-\frac{8H\dot{\eta}}{F}=\frac{\dot{\phi^2}}{6H^2F}+\frac{V}{3H^2F}-\frac{\dot{F}}{HF}+\frac{\rho_m}{3H^2F}
\ee
which allows us to define the following dynamical variables
\be\label{eq7}
\begin{aligned}
&&x=\frac{\dot{\phi}^2}{6H^2F},\;\;\; y=\frac{V}{3H^2F},\;\;\; f=\frac{\dot{F}}{HF}\\
&& g=\frac{8H\dot{\eta}}{F},\;\;\; \Omega_m=\frac{\rho_m}{3H^2F},\;\;\; \epsilon=\frac{\dot{H}}{H^2}
\end{aligned}
\ee
In terms of the variables (\ref{eq7}) the Friedmann equation (\ref{eq3}) becomes the restriction
\be\label{eq8}
1=x+y-f+g+\Omega_m
\ee
Note that due to the interaction term in the denominator, the density parameters $\Omega_m$ and $\Omega_{\phi}$ should be interpreted as effective density parameters, where we define $\Omega_{\phi}=x+y-f+g$. Using the slow-roll variable $N=\ln{a}$ and taking the derivatives with respect to $N$ one finds
\be\label{eq9}
f'=\frac{1}{H}\frac{df}{dt}=\frac{1}{H}\left[\frac{\ddot{F}}{HF}-\frac{\dot{F}\dot{H}}{H^2F}-\frac{\dot{F}^2}{HF^2}\right]=\frac{\ddot{F}}{H^2F}-f\epsilon-f^2
\ee
\be\label{eq10}
g'=\frac{1}{H}\left[\frac{8\dot{H}\dot{\eta}}{F}+\frac{8H\ddot{\eta}}{F}-\frac{8H\dot{F}\dot{\eta}}{F^2}\right]=\frac{8\ddot{\eta}}{F}+g\epsilon-gf
\ee
where " $'$ " means the derivative with respect to $N$. From the Eq. (\ref{eq4}) and using (\ref{eq9}) and (\ref{eq10}) follows
\be\label{eq11}
2\epsilon(1-g)=-6x-(f'+f\epsilon+f^2)+f+(g'-g\epsilon+gf)-g-3(1+w_m)\Omega_m
\ee
Note that the matter density parameter $\Omega_m$ can be replaced from the Eq. (\ref{eq8}) into Eq. (\ref{eq11}), giving the equation
\be\label{eq11a}
2\epsilon(1-g)=-6x-(f'+f\epsilon+f^2)+f+(g'-g\epsilon+gf)-g-3(1+w_m)(1-x-y+f-g)
\ee
For the variables $x$ and $y$ it follows
\be\label{eq12}
x'=\frac{1}{H}\left[\frac{\dot{\phi}\ddot{\phi}}{3H^2F}-\frac{\dot{H}\dot{\phi}^2}{3H^3F}-\frac{\dot{\phi}^2\dot{F}}{6H^2F^2}\right]=\frac{\dot{\phi}\ddot{\phi}}{3H^3F}-2x\epsilon -x f
\ee
\be\label{eq13}
y'=\frac{1}{H}\left[\frac{\dot{V}}{3H^2F}-\frac{2V\dot{H}}{3H^3F}-\frac{V\dot{F}}{3H^2F^2}\right]=\frac{\dot{V}}{3H^3F}-2y\epsilon-y f
\ee
Multiplying the equation of motion (\ref{eq5}) by $\dot{\phi}$ and using the product $\dot{\phi}\ddot{\phi}$ from (\ref{eq12}) one finds
\be\label{eq14}
x'+2x\epsilon+xf+6x+y'+2y\epsilon+yf-f(2+\epsilon)+g(1+\epsilon)=0
\ee
In order to deal with the derivative of the potential and to complete the autonomous system we define the three parameters $b$, $c$ and $d$ as follows 
\be\label{eq15}
b=\frac{1}{dF/d\phi}\frac{d^2F}{d\phi^2}\phi,\;\;\; c=\frac{1}{V}\frac{dV}{d\phi}\phi,\;\;\; d=\frac{1}{d\eta/d\phi}\frac{d^2\eta}{d\phi^2}\phi
\ee
These parameters are related to the potential and the couplings, characterizing the main properties of the model. In what follows we restrict the model to the case when the parameters $b$, $c$ and $d$ are constant, which imply restrictions on the functional form of the couplings and potential.  Additionally, we introduce the new dynamical variable $\Gamma$:
\be\label{eq16}
\Gamma=\frac{1}{F}\frac{dF}{d\phi}\phi
\ee
using the constant parameters $b$, $c$, $d$ and the variable $\Gamma$, the dynamical equations for the variables $y, f, g, \Gamma$ can be reduced to
\be\label{eq17}
y'=\frac{c}{\Gamma}f y-2y\epsilon-y f
\ee
\be\label{eq18}
f'=\frac{b}{\Gamma}f^2+\frac{1}{2}f\frac{x'}{x}-\frac{1}{2}f^2
\ee
\be\label{eq19}
g'=2\epsilon g+\frac{d}{\Gamma}g f+\frac{1}{2}g\frac{x'}{x}-\frac{1}{2}gf
\ee
\be\label{eq20}
\Gamma'=b f+f-\Gamma f.
\ee
The equations  (\ref{eq11a}) and (\ref{eq14}) together with the equations (\ref{eq17})-(\ref{eq20}) form the autonomous system. Here we took into account that, after using the restriction (\ref{eq8}), the Eq. (\ref{eq11}) takes the form of Eq. (\ref{eq11a}). 
%%%%%%%%%%%%%%%%%%%%%%%%%%%%%%%%%%%%%%%%%%%%%%%%%%%%%%%%%%%%%%
%%%%%%%%%%%%%%%%%%%%%%%%%%%%%%%%%%%%%%%%%%%%%%%%%%%%%%%%%%%%%%
\section{The critical points}
The explicit expressions for $x',y',f',g',\Gamma'$ and $\epsilon$ are found by solving the simultaneous system of equations  (\ref{eq11a}),  (\ref{eq14}), (\ref{eq17})-(\ref{eq20}), and are given by
\be\label{eq21}
\begin{aligned}
x'=&-\frac{1}{D}\Big[x (2 f (b f (f - g - 2 x) + d g (-f + g + 2 x) + c (2 + f - 3 g) y) + \\
&\Gamma (f^3 - 2 f^2 (g - 3 w) + 2 (g^2 (-1 + 3 w) + 6 x (1 - x + y + w (-1 + x + y)) + \\
 &g (-1 - 17 x + 3 y + 3 w (-1 + 3 x + y))) +
 f (g^2 - 2 g (-3 + 6 w + 2 x) - \\ &2 (1 - 7 x + 3 y + 3 w (-1 + 3 x + y)))))\Big],
 \end{aligned}
\ee
\be\label{eq22}
\begin{aligned}
y'=& \frac{1}{D}\Big[y (f (4 (b f - d g) x + c (f^2 + g^2 + 4 x - 4 g x + 2 g y - 2 f (g + y))) - \\
& \Gamma (f^3 - 2 f^2 (2 + g) + f (g^2 + g (6 - 4 x) + 4 (2 - 3 w) x) - \\ &  2 (g^2 + g (2 - 6 w) x + 6 x (1 + x - y - w (-1 + x + y)))))\Big],
\end{aligned}
\ee
\be\label{eq23}
\begin{aligned}
f'=&-\frac{1}{D}\Big[f (f (d g (-f + g + 2 x) + b (-g^2 + f (g - 2 x) - 4 x + 4 g x) + \\
 &c (2 + f - 3 g) y) + \Gamma (f^3 + g^2 (-1 + 3 w) + f^2 (-2 g + 3 w) + \\
 & 6 x (1 - x + y + w (-1 + x + y)) +  f (-1 + g^2 + g (3 - 6 w - 4 x) + 9 x - 3 y - \\
 & 3 w (-1 + 3 x + y)) +  g (-1 - 17 x + 3 y + 3 w (-1 + 3 x + y))))\Big],
 \end{aligned}
\ee
\be\label{eq24}
\begin{aligned}
g'=&-\frac{1}{D}\Big[g (f (b f (f - g + 2 x) + d (-f^2 + f g + 2 (-2 + g) x) - \\
  & c (-2 + f + g) y) + \Gamma (f^3 + g^2 (1 + 3 w) + f^2 (4 - 2 g + 3 w) - \\
  & g (1 + 13 x + 3 w (1 + x - y) - 3 y) +  f (-1 + g^2 + 5 x - g (3 + 6 w + 4 x) + \\ & 3 w (1 + x - y) - 
         3 y) - 6 x (-3 - x + y + w (-1 + x + y))))\Big],
\end{aligned}
\ee
\be\label{eq25}
\Gamma'=\left(1-\Gamma+b\right)f,
\ee
\be\label{eq26}
\begin{aligned}
\epsilon=&\frac{1}{D}\Big[f \left(-2 b f x + 2 d g x + c (f - g) y\right) - \Gamma (2 f^2 + g^2 + g (2 - 6 w) x +\\ & f (-3 g - 2 x + 6 w x) + 6 x (1 + x - y - w (-1 + x + y)))\Big],
\end{aligned}
\ee
where
$$D=\Gamma \left(f^2 - 2 f g + g^2 + 4 x - 4 g x\right).$$
The equation for $\epsilon$ gives the effective equation of state as $w_{eff}=-1-2\epsilon/3$.
In order to solve this system we need to specify the model, and what we will do is to impose restrictions on the parameters $b$, $c$ and $d$ in the following two cases. \\
%%%%%%%%%%%%%%%%%%%%%%%%%%%%%%%%%%%%%%%%%%%%%%%%%%%%%%%
\noindent {\bf  1. Power-law couplings and potential}\\
It is necessary to annotate that the high dimensionality of the phase space prevents an effective
graphical description of the phase space, and therefore we will limit ourselves to give analytical considerations, and to illustrate some results in two dimensional projections. 
From (\ref{eq15}) and taking into account that the parameters $b$, $c$ and $d$ are constants,  we find the power-law behavior
\be\label{eq27}
h(\phi)\propto \phi^{b+1},\;\;\; V(\phi)\propto \phi^c,\;\;\; \eta(\phi)\propto \phi^{d+1}
\ee
where we used $F(\phi)=1/\kappa^2-h(\phi)$ and $b$, $c$ and $d$ are in general real numbers, but we restrict them to integers. In fact the restrictions (\ref{eq15}) were considered keeping in mind the power-law behavior for the couplings and potential (see \cite{msami}). The critical points for the system satisfying the equations $x'=0, y'=0, f'=0, g'=0, \Gamma'=0$ are listed below, where the stability of the fixed points is determined by evaluating the eigenvalues of the Hessian matrix associated with the system. After solving the equations for the critical points we find.\\
\noindent
{\bf A1}:  $(x,y,f,g,\Gamma)=(1,0,0,0,1+b)$. This point is dominated by the kinetic energy of the scalar field, where $w_{eff}=1$, $\Omega_{\phi}=1$ and $\Omega_m=0$. This is unstable critical point with eigenvalues $\left[-6, 6, 0, 0, 3 (1 - w_m)\right]$.\\
\noindent {\bf A2}: $(x,y,f,g)=(0,0,0,1)$. This point is dominated by the Gauss-Bonnet coupling with $\Omega_{\phi}=1$, and the corresponding effective EoS, $w_{eff}=-1/3$, lies in the deceleration-acceleration divide. The eigenvalues are $\left[4, 2, 2, 0, -1 - 3 w_m\right]$ and the point is saddle (we assume $0\le w_m\le 1$).\\
\noindent{\bf A3}: $(x,y,f,g,\Gamma)=(-1/5,0,0,6/5,1+b)$. This fixed point is dominated by the scalar field ($\Omega_{\phi}=1$) and is a de Sitter solution with $w_{eff}=-1$. The negative sign of $x$ indicates phantom behavior and the eigenvalues $\left[0,0,0,-3,-3(1+w_m)\right]$ indicate that at least the point is saddle. The three zero eigenvalues make difficult to analyze the stability, but since the rest of the eigenvalues are negative, we can say that the stability is marginal.
This solution could correspond to an unstable inflationary phase which evolves towards a matter or dark energy dominated phase.\\
%%%%%%%%%%%%%%%%%%%%%%%%%%%%%%%%%%%%%%%%%%%%%%%%%%%%%%%%%%%%
\noindent {\bf A4}: $(x,y,f,g,\Gamma)=(0,0,-1,0,1+b)$.  The eigenvalues are  $[\frac{-1 + b}{1 + b}, 1, \frac{5 + 5 b - c}{1 + b}, \frac{-4 - 3 b - d}{1 + b}, 2 - 3 w_m]$. This point is controlled by the non-minimal coupling ($\Omega_{\phi}=1$) and gives a solution that leads to an equation of state corresponding to radiation $w_{eff}=1/3$. At this critical point the potential and the GB coupling disappear, and is a saddle point depending on the values of the parameters $b,c,d$ and $w_m$. Thus for instance, if $-1<b<1$, $c>5(1+b)$, $d>-4-3b$ and $w>2/3$ all the eigenvalues except one are negative. For background radiation ($w_m=1/3$) or dust matter ($w_m=0$) three of the eigenvalues might take negative values. In the case of background matter given by radiation, this critical point presents a scaling behavior. At this point, despite the presence of the background matter in form of radiation or dust, the universe becomes radiation dominated, but due to the saddle character, this point could represent a transient phase of radiation dominated universe.\\ 
%%%%%%%%%%%%%%%%%%%%%%%%%%%%%%%%%%%%%%%%%%%%%%%%%%%%%%%%%%%%
\noindent {\bf A5}: $(x,y,f,g,\Gamma)=(0,\frac{5 + 5 b - c}{1 + b + c},\frac{4 + 4 b - 2 c}{1 + b + c},0,1+b)$. This critical point is dominated by the potential and the non-minimal coupling with 
\be\label{eq21}
w_{eff}=-1+\frac{2 (1 + b - c) (2 + 2 b - c)}{3 (1 + b) (1 + b + c)},
\ee
and $\Omega_{\phi}=1$. The effective EoS describes different regimes depending on the values of $b,c,d$. Note that for the scalar field dominated universe the effective EoS $w_{eff}$ and the dark energy EoS $w_{DE}$ take the same value. The eigenvalues are given by
$$
\begin{aligned}
&\Big[-\frac{2 (-1 + b) (2 + 2 b - c)}{(1 + b) (1 + b + c)},\frac{-4 - 4 b + 2 c}{1 + b + c},\frac{-5 - 5 b + c}{1 + b},-\frac{2 (2 + 2 b - c) (1 + 2 b - c - d)}{(1 + b) (1 + b + c)},\\
& -\frac{3 + 3 b^2 + 7 c - 2 c^2 + b (6 + 7 c)+3w_m(1+c+b^2+2b+bc)}{(1 + b) (1 + b + c)}\Big].
\end{aligned}
$$
In the case $c=1+b$ we obtain the de Sitter solution with $w_{eff}=-1$, with eigenvalues given by $$\left[\frac{1 - b}{1 + b}, -1, -4, \frac{-b + d}{1 + b}, -4-3w_m\right].$$This solution is a stable fixed point for any type of matter with $0\le w_m\le1$, whenever $b>1$ and $d<b$ or $b<-1$ and $d>b$. The de Sitter solution for the quadratic potential, corresponding to $c=2$ ($V\propto \phi^2$) ($b=1$, $h\propto \phi^2$), has eigenvalues $[0,-1,-4,\frac{1}{2}(-1+d),-4-3w_m]$ and is marginally stable since four eigenvalues are negative (whenever $d<1$) and there is only one zero eigenvalue, but numerical study shows that the point is an attractor as can be seen in Fig. 1. The Higgs-type potential ($V\propto \phi^4$) is obtained for $c=4$ ($b=3$, $h\propto \phi^4$) and leads to de Sitter stable solution whenever $d<3$. The cubic non-minimal coupling, $h\propto \phi^3$, and cubic potential $V\propto \phi^3$, also give stable de Sitter solution with eigenvalues $[-(1/3), -1, -4, 1/3 (-2 + d), -4]$, for any $d<2$. The de Sitter solution can also be obtained for $c=2+2b$ with the eigenvalues $[0,0,-3,0,-3(1+w)]$, which include the standard non-minimal coupling ($b=1$, $h\propto\phi^2$) and the Higgs-type potential $V\propto\phi^4$. In this case the point is at least a saddle point, but it is difficult to analyze the stability because of the the three zero eigenvalues. In Figs. 1 and 2 we show the behavior of some trajectories around the point {\bf A5} for $b=1$, $c=2$ and $b=1$, $c=4$ respectively.
We can also consider values in a region around $w_{eff}=-1$, which are consistent with observations. Thus, the values $b=4, c=4$, give $w_{eff}\approx -0.91$ and the critical point $(0,7/3,4/3,0,5)$ is stable with eigenvalues  $[-4/5, -4/3, -21/5, -16/15, -61/15]$ (taking $d=1$, $\eta\propto \phi^2$). The critical point $(0,11/7,4/7,0,3)$ with eigenvalues $[-4/21, -4/7, -11/3, -4/21, -79/21]$, corresponding to $b=2,c=4$, gives stable phantom solution with $w_{eff}\approx -1.06$ (taking $d=0$, $\eta\propto \phi$). 
In fact the general conditions for the existence of stable quintessence fixed point, assuming $0\le w_m\le 1$, are $b<-1$, $1+b< c< (3-\sqrt{10})(1+b)$ and $d> 1+2b-c$ or $b>1$, $(3-\sqrt{10})(1+b)< c< 1+b$ and $d<1+2b-c$, and the general conditions for the existence of stable phantom fixed point are $b<-1$, $2+2b<c< 1+b$ and $d>1+2b-c$ or $b>1$, $1+b< c<2+2b$ and $d<1+2b-c$, for $w_m$ in the interval $0\le w_m\le 1$.
This point has all the necessary properties for the description of late time cosmological scenarios.\\
\begin{center}
\includegraphics [scale=0.6]{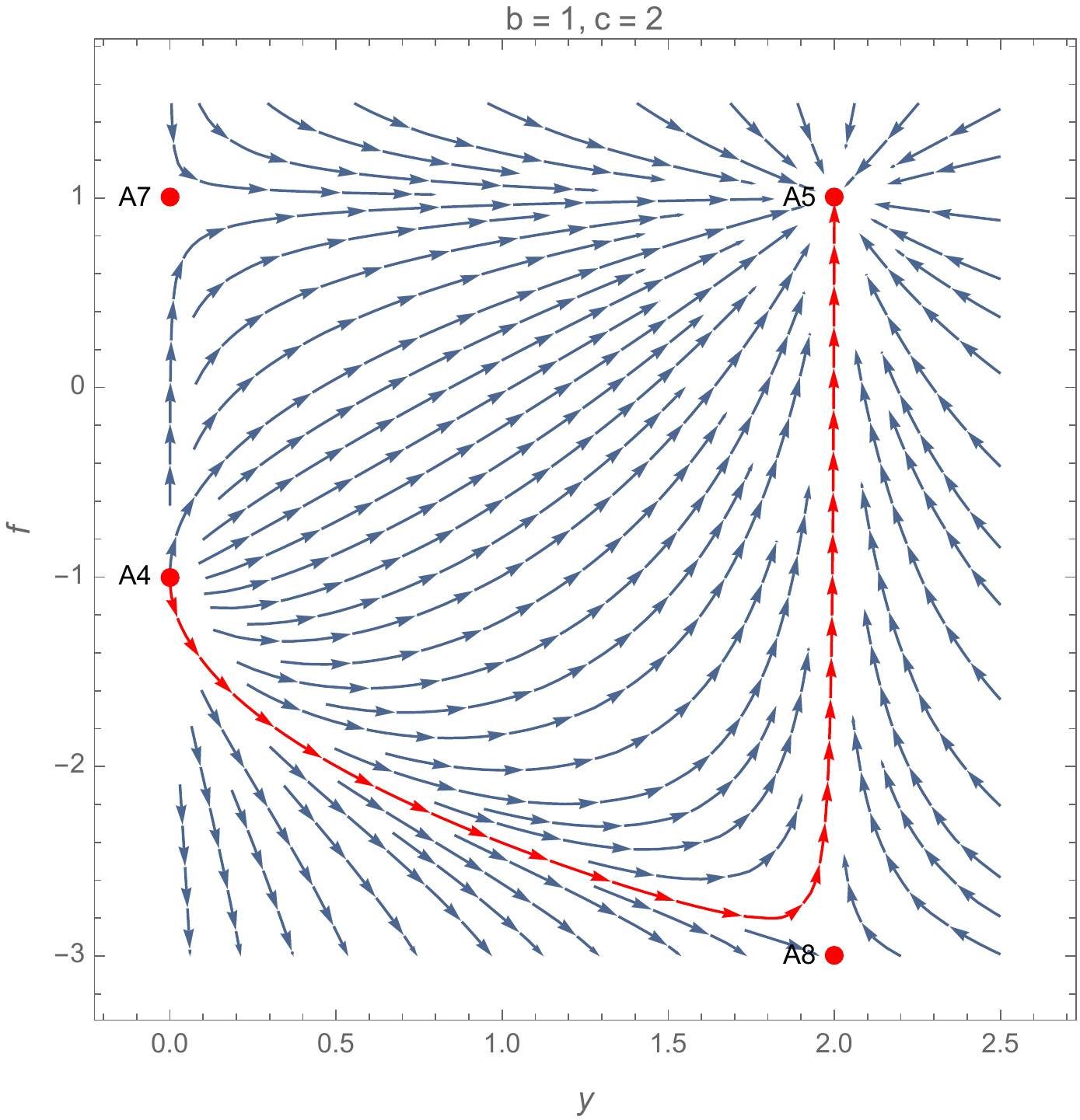}
\end{center} 
\begin{center}
{Fig. 1 \it The projection of the phase portrait of the model on the $yf$-plane for the standard non-minimal coupling $h\propto\phi^2$ and the quadratic potential $V\propto\phi^2$ ($b=1$, $c=2$), taking $w_m=0$. The graphic shows that the de Sitter solution for the point {\bf A5} behaves as an attractor on the $yf$-plane, the point {\bf A4} (radiation dominated universe with $\Omega_{\phi}=1$) is unstable on this plane and {\bf A7} (which is not physical in this case since $\Omega_m=2$) behaves as saddle. The only negative eigenvalue of {\bf A4} is located on the $g$-axis. The de Sitter solution for the point {\bf A5} in the case  $b>1$ and $d<b$ is an attractor and could correspond to a final stage of vacuum dominated universe.}
\end{center}
%%%%%%%%%%%%%%%%%%%%%%%%%%%%%%%%%%%%
\begin{center}
\includegraphics [scale=0.6]{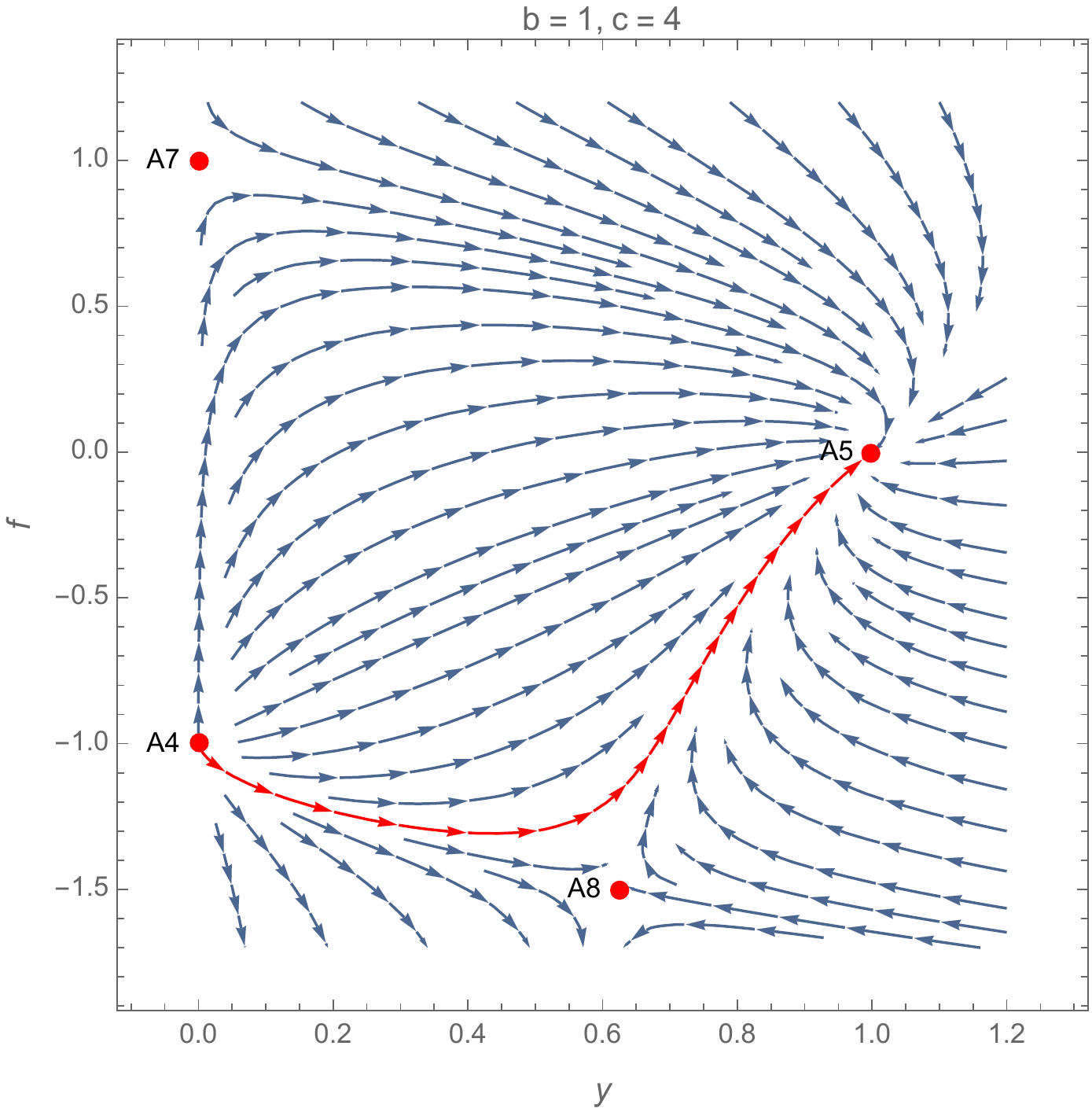}
\end{center} 
\begin{center}
{Fig. 2 \it The projection of the phase portrait of the model on the $yf$-plane for the standard non-minimal coupling $h\propto\phi^2$ and the Higgs-type potential $V\propto\phi^4$ ($b=1$, $c=4$), with $w_m=0$. The de Sitter solution for the point {\bf A5} also shows an attractor character on the plane $yf$. The points {\bf A4} and {\bf A7} present a behavior similar to the case $c=2$.}
\end{center}
%\begin{figure}[hbtp]
%\includegraphics [scale=0.5]{plane-yf.pdf}
%\includegraphics [scale=0.5]{plane-yf-Higgs.pdf}
%\caption{The projection of the phase portrait of the model on the $yf$-plane for the standard non-minimal coupling $h\propto\phi^2$ and the quadratic potential $V\propto\phi^2$ ($b=1$, $c=2$), taking $w_m=0$. The graphic shows that the de Sitter solution for the point {\bf A5} behaves as an attractor and the points {\bf A4} (radiation dominated universe with $\Omega_{\phi}=1$) and {\bf A7} (radiation dominated universe with $\Omega_m=1$) behave as saddle. The only negative eigenvalue of {\bf A4} is located on the $g$-axis. The point {\bf A5} could correspond to a final stage of vacuum dominated universe.} 
%\end{figure}
%%%%%%%%%%%%%%%%%%%%%%%%%%%%%%%%%%%%%%%%%%%%%%%%%%%%%%%%%%%%%%
\noindent
{\bf  A6}: $(x,y,f,g,\Gamma)=(0,0,\frac{2(1+b)}{2+b+d},\frac{4 + 3 b + d}{2 + b + d},1+b)$. This critical point is dominated by the non-minimal and GB couplings, with $\Omega_{\phi}=1$ and the effective EoS given by
\be\label{eq22}
w_{eff}=-1-\frac{2 (b - d)}{3 (2 + b + d)}.
\ee
This equation gives the three possible accelerating regimes for the late time Universe: quintessence phase with $w_{eff}>-1$ for $d>b$, de Sitter Universe with $w_{eff}=-1$ for $b=d$ and the phantom phase with $w_{eff}<-1$ for $d<b$. The stability properties of this point can be deduced from the corresponding eigenvalues given by
$$
\begin{aligned}
&\Big[\frac{2(1-b)}{2+b+d},-\frac{2(1+b)}{2+b+d},-\frac{2(1+2b-c-d}{2+b+d},-\frac{2(4+3b+d)}{2+b+d},\\
&-\frac{8+7b+d+3w_m(2+b+d)}{2+b+d}\Big].
\end{aligned}
$$
The eigenvalues for the de Sitter solution, which is obtained for $d=b$, reduce to
$$
\left[\frac{1-b}{1+b},-1,-\frac{1+b-c}{1+b},-4,-4-3w_m\right],
$$
indicating that the de Sitter fixed point is an attractor whenever $b>1$ and $c<1+b$, or $b<-1$ and $c>1+b$, for any type of matter with $0\le w_m \le 1$. This includes constant potential $V=cons.$ ($c=0$), quadratic potential $V\propto \phi^2$ ($c=2$) and fourth order potential $V\propto \phi^4$ ($c=4$). The case $b=1$, which leads to the standard $\phi^2$ non-minimal coupling, is a marginally stable fixed point with eigenvalues $[0,-1,-1+c/2,-4,-4-3w_m]$. In Fig. 3 we illustrate the behavior of the system around the point {\bf A6} for the de Sitter solution with non-minimal coupling $h\propto \phi^2$ and the GB coupling $\eta\propto \phi^2$. To analyze the properties of stability of the quintessence or phantom fixed points we consider the matter EoS in the interval $0\le w_m\le 1$. In this case, the condition of stability for the fixed point in the quintessence phase reduces to $b<-1$, $c>1+b$ and $1+2b-c<d\le b$ or $b>1$, $c<1+b$ and $b\le d<1+2b-c$,
and any phantom fixed point is stable if one of the following sets of inequalities: $b<-1$, $3(1+b)<c\le 1+b$ and $1+2b-c<d<-2-b$ or $b<-1$, $c>1+b$ and $b\le d<-2-b$ or $b>1$, $c<1+b$ and $-2-b<d\le b$ or $b>1$, $1+b\le c<3(1+b)$ and $-2-b<d<1+2b-c$
is satisfied. Thus for instance, the values $b=3$, $d=2$ give the phantom fixed point with $w_{eff}\approx -1.095$ and the corresponding eigenvalue $[-4/7,-8/7,-2(5-c)/7,-30/7,-31/7]$, indicating that the stability depends on the potential and the solution is stable for $V\propto \phi^c$, $c=0,1,...,4$. A quintessence fixed point with $w_{eff}\approx -0.93$ is obtained for $b=3$ and $d=4$, with eigenvalues $[-4/9,-8/9,-2(3-c)/9,-34/9,-11/3]$. Particularly, the cosmological constant ($c=0$) and the quadratic potential ($c=2$) give quintessence attractor.\\
\begin{center}
\includegraphics [scale=0.6]{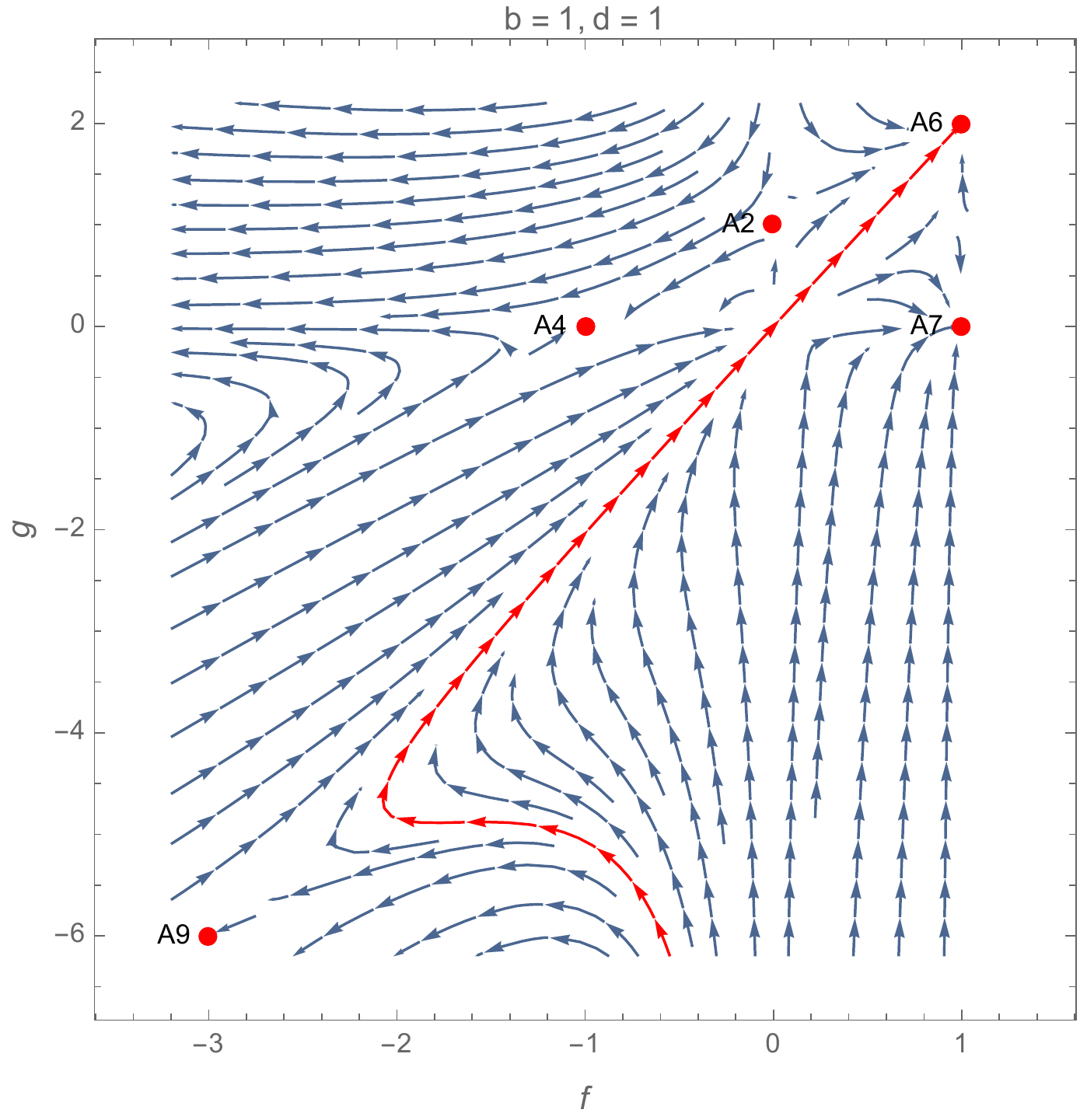}
\end{center} 
\begin{center}
{Fig. 3 \it The projection of the phase portrait of the model on the $fg$-plane for the standard non-minimal coupling $h\propto\phi^2$ and the GB coupling $\eta\propto\phi^2$ ($b=1$, $d=1$), with $w_m=0$. The point {\bf A2} is unstable on this plane and corresponds to the transition between decelerated and accelerated regimes.  The de Sitter solution for the point {\bf A5} is stable on this plane, and in the case $b>1$ and $c<1+b$, is an attractor that could describe the final stage of vacuum dominated universe.}
\end{center}
%%%%%%%%%%%%%%%%%%%%%%%%%%%%%%%%%%%%%%%%%%%%%%%%%%%%%%%%%%%%%%
\noindent
{\bf  A7}: $(x,y,f,g,\Gamma)=(0,0,1-3w_m,0,1+b)$ with eigenvalues
$$
\begin{aligned}
&\Big[\frac{(-1+b)(-1+3w_m)}{1+b},-1+3w_m,-2+3w_m,\frac{3+c+3w_m(1-c)+3b(1+w_m)}{1+b},\\
&\frac{6-5b-30w_m+3bw_m}{1+b}\Big].
\end{aligned}
$$
To this fixed point the matter and the non-minimal coupling contribute giving $w_{eff}=1/3$ with $\Omega_{\phi}=-1+3w_m$ and $\Omega_m=2-3w_m$. The positivity of the density parameters $\Omega_m$ and $\Omega_{\phi}$ impose the restriction $1/3\le w_m\le 2/3$, which excludes the pressureless dust matter. If the background matter consists of radiation ($w_m=1/3$), then the fixed point becomes a scaling solution and the universe becomes radiation-dominated with $\Omega_{\phi}=0$ and $\Omega_m=1$. At this saddle point with eigenvalues $[0,0,-1,4,-4]$, which do not depend on $b$, the system can reach the conformal invariance and can be considered as a transient phase of radiation dominated universe.\\
%%%%%%%%%%%%%%%%%%%%%%%%%%%%%%%%%%%%%%%%%%%%%%%%%%%%%%%%%%%%%
\noindent
{\bf  A8}: $(x,y,f,g,\Gamma)=\left(0,\frac{(1+b)(3+c+3w_m(1-c)+3b(1+w_m))}{2c^2},-\frac{3(1+b)(1+w_m)}{c},0,1+b\right)$. The scalar field density parameter is $\Omega_{\phi}=\frac{(1+b)(3+7c+3w_m(1+c)+3b(1+w_m)}{2c^2}$ and the EoS is $w_{eff}=-1-\frac{(1+b-c)(1+w_m)}{c}$ which gives de Sitter solution for $c=1+b$ with eigenvalues $[\frac{3 (-1 + b) (1 + w_m)}{1 + b}, 3 (1 + w_m), \frac{3 (b - d) (1 + w_m)}{1 + b}, -4, 4 + 3 w_m]$, showing that this is a saddle point with at least two positive eigenvalues.
Though this point has quintessence ( for $c>1+b$) and phantom (for  for $c<1+b$) solutions, it was found that there are not integer 
values for the parameters that simultaneously satisfy the restriction $0\le\Omega_{\phi}\le 1$ and give adequate values to  $w_{eff}$
(i.e.  $w_{eff}$ is out of the region of physical interest) for $0\le w_m\le 1$.\\
%%%%%%%%%%%%%%%%%%%%%%%%%%%%%%%%%%%%%%%%%%%%%%%%%%%%%%%%%%%%%
\noindent
{\bf  A9}: $(x,y,f,g,\Gamma)=\left(0,0,-\frac{3(1+b)(1+3w_m)}{1+2b-d},-\frac{3(1+b)(1+w_m)(-4+d(1-3w_m)+b(-5+3w_m))}{(1+2b-d)(-2+b(-1+3w_m)-d(1+3w_m))},1+b\right)$. The density parameter of the scalar field is $\Omega_{\phi}=\frac{6(1+b)(1+w_m)}{-2+b(-1+3w_m)-d(1+3w_m)}$ and the corresponding effective EoS is $w_{eff}=-1+\frac{(b-d)(1+w_m)}{1+2b-d}$, which  leads to de Sitter with eigenvalues $[\frac{3 (-1 + b)(1+w_m)}{1 + b}, 3(1+w_m), \frac{3 (1 + b - c)(1+w_m)}{1 + b},4+3w_m,-4]$, showing that this point is saddle, but this point is not physical since $\Omega_{\phi}$ takes negative values for $0\le w_m\le 1$. As in the previous point, the conditions $0\le \Omega_{\phi}\le 1$ and physically meaningful values of  $w_{eff}$ can not be reached simultaneously with adequate integer values of the parameters.\\
%%%%%%%%%%%%%%%%%%%%%%%%%%%%%%%%%%%%%%%%%%%%%%%%%%%%%%%%%%%%%
%%%%%%%%%%%%%%%%%%%%%%%%%%%%%%%%%%%%%%%%%%%%%%%%%%%%%%%%%%%%%
\noindent {\bf  2. Exponential function for couplings and potential}\\
In this case we introduce the following restrictions on the couplings and potential by defining the constant parameters $b$, $c$, and $d$ as
\be\label{eq30}
b=\frac{1}{dF/d\phi}\frac{d^2F}{d\phi^2},\;\;\; c=\frac{1}{V}\frac{dV}{d\phi},\;\;\; d=\frac{1}{d\eta/d\phi}\frac{d^2\eta}{d\phi^2}.
\ee
And the new dynamical variable $\Gamma$ is defined now as
\be\label{eq30a}
\Gamma=\frac{1}{F}\frac{dF}{d\phi}
\ee
Integrating the equations (\ref{eq30}) with respect to the scalar field, one finds
\be\label{eq31}
h(\phi)\propto e^{b\phi},\;\;\; V(\phi)\propto e^{c\phi},\;\;\; \eta(\phi)\propto e^{d\phi}
\ee
where $b$, $c$ and $d$ are real numbers. The only equation of the autonomous system (\ref{eq21})-(\ref{eq25}) that changes is the one related with the variable $\Gamma$ which reduces to
\be\label{eq32}
\Gamma'=\left(-\Gamma+b\right)f.
\ee
The critical points of the system that are given by {\bf  B1}= $(1,0,0,0,b)$, {\bf  B2}=$(0,0,0,1)$, {\bf  B3}=$(-1/5,0,0,6/5,b)$ and {\bf  B4}=$(0,0,-1,0,b)$, have the same stability properties and lead to the same $w_{eff}$ and $\Omega_{\phi}$ as the points {\bf  A1}, {\bf  A2}, {\bf  A3} and {\bf  A4} respectively. Other critical points are the following.\\
\noindent
{\bf B5}:  $(x,y,f,g,\Gamma)=(0,\frac{5b-c}{b+c},\frac{4b-2c}{b+c},0,b)$. This fixed point is dominated by the scalar field, specifically by the potential and non-minimal coupling, with $\Omega_{\phi}=1$, leading to the effective EoS
\be\label{eq33}
w_{eff}=-1+\frac{2(b-c)(2b-c)}{3b(b+c)}
\ee
with eigenvalues
$$
\begin{aligned}
&\Big[-\frac{2(2b-c)}{b+c},-\frac{2(2b-c)}{b+c},-\frac{10b^2-7bc+c^2}{b(2b-c)},-\frac{2(2b-c)(2b-c-d)}{b(b+c)},\\
& \frac{-6b^4-17b^3 c+9bc^3-2c^4-6b^4 w_m-9b^3 cw_m+3bc^3 w_m}{b(2b-c)(b+c)^2}\Big]
\end{aligned}
$$
The scaling solution with $w_{eff}=w_m$, from to the Eq. (\ref{eq33}), takes place if 
\be\label{eq33a}
c=\frac{1}{4}\left(9b+3bw_m-b\sqrt{73+78w_m+9w_m^2}\right)
\ee
replacing this restriction for $c$ in the eigenvalues we find that the scaling solution corresponding to this critical point is stable for $0\le w_m\le 1$ if the following conditions are satisfied: $b<0$ and $d>\frac{1}{4}(-b-3bw_m-\sqrt{73b^2+78b^2w_m+9b^2w_m^2})$ or $b>0$ and $d<\frac{1}{4}(-b-3bw_m+\sqrt{73b^2+78b^2w_m+9b^2w_m^2})$. So, if we define the potential so that the potential parameter $c$ depends on the non-minimal coupling parameter $b$ and $w_m$ as given by the equation (\ref{eq33a}), then the critical point is a scaling attractor if the above inequalities are satisfied. This result provides a cosmological scenario where the energy density of the scalar field behaves similarly to the background fluid in either the radiation or matter era, but with the dominance of the scalar field.\\
\noindent According to the equation (\ref{eq33}) the de Sitter solution takes place for $c=b$ and $c=2b$. In the case $c=b$ the eigenvalues reduce to $[-1,-1,-4,-1+\frac{d}{b},-4-3w]$, indicating that the de Sitter solution is a stable node (attractor) for any type of matter with $0\le w_m \le 1$ and for $d<b$, and is a saddle point if $d>b$. As follows from the expression for the eigenvalues, the case $c=2b$ leads to zero and indeterminate eigenvalues and therefore can not be considered. On the other hand, the quintessence behavior ($w_{eff}>-1$) takes place for the restriction $\frac{2(b-c)(2b-c)}{3b(b+c)}>0$. To analyze the stability in this case, we limit ourselves to the relevant interval $0\le w_m\le 1$, and them according to the expression for the eigenvalues, the quintessence fixed point is an attractor if the inequalities $b<0$, $b\le c\le (3+\sqrt{10})b$ and $d>2b-c$ or $b>0$, $(3-\sqrt{10})b\le c\le b$ and $d<2b-c$, are satisfied. The fixed point describes phantom phase or super accelerated expansion in the case $\frac{2(b-c)(2b-c)}{3b(b+c)}<0$. This phase is stable if the parameters satisfy one of the following sets of inequalities $b<0$, $2b<c\le b$ and $d>2b-c$ or $b>0$, $b\le c<2b$ and $d<2b-c$. In the quintessence and phantom phases the effective EoS $w_{eff}$ can be as close to $-1$ as we need, since the parameters $b$, $c$ and $d$ are real numbers. So, this new fixed point is very interesting cosmological solution since it can account for the accelerating universe.\\
%%%%%%%%%%%%%%%%%%%%%%%%%%%%%%%%%%%%%%%%%%%%%%%%%%%%%%%%%%
\noindent
{\bf B6}:  $(x,y,f,g,\Gamma)=(0,0,\frac{2b}{b+d},\frac{3b+d}{b+d},b)$. This fixed point dominated by the scalar field (non-minimal and GB couplings), with $\Omega_{\phi}=1$, leads to the effective EoS
\be\label{eq34}
w_{eff}=-1-\frac{2(b-d)}{3(b+d)}
\ee
with eigenvalues
$$
\left[-\frac{2b}{b+d},-\frac{2b}{b+d},-\frac{2(2b-c-d)}{b+d},-\frac{2(3b+d)}{b+d},-\frac{7b+d+3w_m(b+d)}{b+d}\right].
$$
The scaling behavior ($w_{eff}=w_m$) takes place if the GB parameter $d$ is related to the non-minimal coupling parameter $b$ as follows: $$d=-\frac{5b+3bw_m}{1+3w_m}.$$ Replacing this expression for $d$ one finds the eigenvalues
$$
\left[\frac{1}{2}(1+3w_m), \frac{1}{2}(1+3w_m), \frac{7b-c+3w_m(3b-c)}{2b},-1+3w_m,\frac{1}{2}(1+3w_m)\right],
$$
which indicates that this critical point, with $w_{eff}=w_m$, is unstable or saddle point for $0\le w_m\le 1$.
For $d=b$ the system reaches a de Sitter fixed point. This point is stable for any $w_m$ in the region $0\le w_m\le 1$ and $c<b$, as follows from the eigenvalues: $[-1,-1,-1+\frac{c}{b},-4,-4-3w_m]$ (if $c>b$ the point is saddle). The EoS also leads to quintessence solutions in the case $\frac{2(b-d)}{3(b+d)}<0$. The quintessence fixed point is stable (assuming $0\le w_m\le 1$) if one of the two sets of inequalities is satisfied:
$b<0$, $c>b$ and $2b-c<d< b$ or $b>0$, $c<b$ and $b< d<2b-c$. The phantom phase is also possible with stable fixed point under one of the following sets of restrictions: $b<0$, $3b<c\le b$ and $2b-c<d<-b$ or $b<0$, $c>b$ and $b<d<-b$ or $b>0$, $c<b$ and $-b<d<b$ or $b>0$, $b\le c<3b$ and $-b<d<2b-c$. As in the point {\bf B5}, in this fixed point the effective EoS can be as close to $-1$ as we want, making of this point an interesting one for the description of the late time Universe.\\

\noindent
{\bf B7}: $(x,y,f,g,\Gamma)=(0,0,1-3w_m,0,b)$. At this fixed point the Universe becomes dominated by the non-minimal coupling and matter with $\Omega_{\phi}=-1+3w_m$ and $\Omega_m=2-3w_m$ which have physical meaning in the region $1/3\le w_m\le 2/3$, excluding the pressureless dust as background matter. The eigenvalues are given by 
$$
\begin{aligned}
&\Big[-1+3w_m,-1+3w_m,-2+3w_m,\frac{c(1-3w_m)+3b(1+w_m)}{b},\\
&\frac{d(1-3w_m)+b(-5+3w_m)}{b}\Big],
\end{aligned}
$$
and the effective EoS  corresponds to radiation $w_{eff}=1/3$.  If the background matter is made up of radiation ($w_m=1/3$), then the fixed point leads to scaling solution and the universe becomes radiation dominated ($\Omega_m=1$). Concerning the stability, this point is saddle with eigenvalues $[0,0,-1,4,-4]$.\\

\noindent
{\bf B8}: $(x,y,f,g,\Gamma)=(0,\frac{b(c-3cw_m+3b(1+w_m))}{2c^2},-\frac{3b(1+w_m)}{c},0,b)$. The effective EoS is given by $w_{eff}=-1-\frac{(b-c)(1+w)}{c}$ and the density parameters are $\Omega_m=\frac{2c^2-3b^2(1+w_m)+bc(7+3w_m)}{2c^2}$ and $\Omega_{\phi}=\frac{b(3b(1+w_m)+c(7+3w_m)}{2c^2}$. The de Sitter solution follows for $c=b$, but the density parameters are out of the physical range for $c=b$. In order to find physical solutions, the density parameters should satisfy the restrictions $0\le\Omega_m\le 1$, $0\le \Omega_{\phi}\le1$ for $w_m$ in the region $0\le w_m \le 1$, but despite the fact that they can be fulfilled, however the effective EoS falls into regions out of cosmological interest.\\

\noindent
{\bf B9}: $(x,y,f,g,\Gamma)=(0,0,-\frac{3b(1+w_m)}{2b-d},-\frac{3b(1+w_m)(d-3dw_m+b(-5+3w_m))}{(2b-d)(b(-1+3w_m)-d(1+3w_m))},b)$. The effective EoS is $w_{eff}=-1+\frac{(b-d)(1+w_m)}{2b-d}$, with density parameters $\Omega_m=\frac{7b+d+3w_m(b+d)}{b+d+3w_m(d-b)}$ and $\Omega_{\phi}=-\frac{6b(1+w_m)}{b+d+3w_m(d-b)}$. As in the previous case, in none of the phases we can obtain all physically meaningful quantities.\\
%%%%%%%%%%%%%%%%%%%%%%%%%%%%%%%%%%%%%%%%%%%%%%%%%%%%%%%%%%
The coordinates of the fixed points allow us to analyze the behavior of the physical quantities. Thus for instance, evaluating $\epsilon$ given in (\ref{eq26}) at the fixed point {\bf A5} one finds from the last equation in (\ref{eq7})
\be\label{eq35}
\dot{H}=-\frac{(1+b-c)(2+2b-c)}{(1+b)(1+b+c)}H^2.
\ee
Integrating this equation gives the power-law solution
\be\label{eq36}
a(t)=a_0(t-t_0)^{\alpha},\;\;\;  \alpha=\frac{(1+b)(1+b+c)}{(1+b-c)(2+2b-c)}.
\ee
Note that for the phantom solution where the power index in (\ref{eq36}) is negative, the scale factor can be written more properly as
\be\label{eq40}
a(t)=\frac{a_0}{(t_c-t)^{|\alpha|}},
\ee
which reflects the Big Rip singularity characteristic of the phantom power-law solutions.\\
From the dynamical variables $f$ and $\Gamma$ defined in (\ref{eq7}) and (\ref{eq16}) evaluated at the fixed point {\bf A5} one finds
\be\label{eq37}
\frac{f}{\Gamma}\Big |_{A5}=\frac{4+4b-2c}{(1+b+c)(1+b)}=\frac{\dot{\phi}}{H\phi}.
\ee
Integrating this equation gives
\be\label{eq38}
\phi=\phi_0(t-t_0)^{\frac{2}{1+b-c}}.
\ee
%For the phantom solution with $c>1+b$ the power index is negative and the scalar field will satisfy the condition $\lim \dot{\phi}%(t)\Big |_{t\rightarrow\infty}=0$, maintaining the consistency with the coordinate $x=0$ for this critical point or, in other words, %conserving the solution in the invariant sub manifold $x=0$. The negative power index also imply that $\lim \phi(t)\Big |_{t%\rightarrow\infty}=0$, indicating that at the fixed point the true Newtonian coupling is restored.
%For the quintessence solution ($c<1+b$), taking the derivative of $\phi$ one finds
%\be\label{eq39}
%\dot{\phi}=\frac{2}{1+b-c}\phi_0(t-t_0)^{\frac{2}{1+b-c}-1}
%\ee
Taking into account the above solutions, the condition $x\rightarrow 0$ at $t\rightarrow \infty$ can be accomplished in general as follows: taking into account that $H\propto t^{-1}$, $h(\phi)\propto \phi^{b+1}$ and $\phi\propto t^{\beta}$ ($\beta=\frac{2}{1+b-c}$) then, if $\beta>0$, at large times we can write for $x$ (see Eq. (\ref{eq7}))
\be\label{eq38a}
x=\frac{\dot{\phi}^2}{6H^2F}\propto \frac{t^2 t^{2(\beta-1)}}{t^{(b+1)\beta}}=\frac{t^{2\beta}}{t^{(b+1)\beta}}
\ee
where we used $h(\phi)\propto \phi^{b+1}$. In order to satisfy the limit $$\lim_{t\to \infty}x=0,$$ the restrictions $\beta>0$, $b>1$ or $\beta<0$, $b<-1$ must be fulfilled. This maintains the consistency with the coordinate $x=0$ for this critical point or, in other words, conserves the solution in the invariant sub manifold $x=0$. Note that for $\beta>0$ (keeping $b>1$), from (\ref{eq38}) it follows that $\lim_{t\to \infty}\phi=\infty$ and this imply, using the expression for the variable $\Gamma$ (using $h(\phi)=\xi \phi^{b+1}$)
\be\label{eq39}
\Gamma=\frac{F'\phi}{F}=\frac{-\xi (b+1)\phi^{b+1}}{\kappa^{-2}-\xi\phi^{b+1}},
\ee
that 
\be\label{eq40}
\lim_{t\to \infty}\Gamma=b+1,
\ee
in complete agreement with the $\Gamma$ coordinate of the critical point {\bf A5}, i. e. $\Gamma\Big |_{A5}=1+b$. In the case of $\beta<0$ ($c>1+b$), we have the limit $\lim_{t\to \infty}\phi=0$ and in order to keep the limit (\ref{eq40}), the parameter $b$ in (\ref{eq39}) must satisfy the condition $b<-1$ (see (\ref{eq38a})). For negative $\beta$, both the scalar field and its time derivative behave asymptotically as $\lim_{t\to \infty}\phi=\lim_{t\to \infty}\dot{\phi}=0$ which imply (whenever $b<-1$), for the $x$-coordinate of the critical point {\bf A5}, that $x=0$. The inequalities $b<-1$ and $c>1+b$ are consistent with the existence of quintessence solutions discussed at the end of the point {\bf A5}, and the restriction $b>1$ is consistent with both, the existence of quintessence and phantom solutions discussed at the end of the point {\bf A5}.\\
%%%%%%%%%%%%%%%%%%%%%%%%%%%%%%%%%%%%%%%%%%%%%%%%%%%%%%%%
\noindent A special attention deserves the de Sitter solution, which for the point {\bf A5} takes place for $c=1+b$ as follows from the expression (\ref{eq21}). This means that $\dot{H}=0$, giving
\be\label{eq41}
H=const=H_0, \;\;\; a(t)=a_0 e^{H_0(t-t_0)}
\ee
and from the relation $f/\Gamma$ at the fixed point, and replacing  $c=1+b$, one finds
\be\label{eq42}
\frac{\dot{\phi}}{H\phi}=\frac{1}{1+b}.
\ee
Integrating this equation gives
\be\label{eq43}
\phi(t)=\phi_0 e^{\frac{H_0}{1+b}(t-t_0)}
\ee
These expressions allow us to analyze the behavior of the coordinate $x$ at the fixed point 
\be\label{eq44}
x\Big |_{A5}=\lim_{t\to \infty}\frac{\dot{\phi}^2}{6H^2F}
\ee
using the expressions (\ref{eq41}) and (\ref{eq43}) for $H$ and $\phi$ we can see that the behavior of $x$ at large times is of the form
\be\label{eq44a}
x\propto e^{(\frac{2}{1+b}-1)H_0(t-t_0)},
\ee 
and we can deduce two possibilities for the scalar field: \\
1) If $b>1$, then $\lim_{t\to \infty}\phi=\infty$ and $\lim_{t\to \infty} x=0$, and\\
2) If $b<-1$, then $\lim_{t\to \infty}\phi=0$ and $\lim_{t\to \infty} x=0$.\\
These behaviors do not affect the coordinate $\Gamma$ of the critical point since the power (b+1) cancels with the denominator in the exponential index of the expression (\ref{eq43}) for the scalar field (see (\ref{eq39})). The effective Newtonian coupling vanishes at $t\to \infty$, independently of the value of $b$. \\
                                                 %%%%%%%%%%%%%%%%%%%%
Concerning the point {\bf A6}, the power-law behavior of the scale factor is given by
\be\label{eq45}
a(t)=a_0(t-t_0)^{\beta}, \;\;\;  \beta=\frac{2+b+d}{d-b},
\ee
and the scalar field satisfies the equation
\be\label{eq46}
\frac{f}{\Gamma}\Big |_{A6}=\frac{2}{2+b+d}=\frac{\dot{\phi}}{H\phi},
\ee
giving
\be\label{eq47}
\phi=\phi_0(t-t_0)^{\frac{2}{d-b}}.
\ee
Applying the equation (\ref{eq44}) to the point {\bf A6} one finds the behavior of $x$
\be\label{eq48}
x\propto \frac{(t-t_0)^{\frac{4}{d-b}}}{\kappa^{-2}-\xi\phi_0^{(1+b)}(t-t_0)^{\frac{2(1+b)}{d-b}}}
\ee
and, assuming $\frac{b+1}{d-b}>0$, then at large $t$ it follows
$$ x\propto t^{\frac{2(1-b)}{d-b}}$$
which lead to two possibilities for the scalar field:\\
1) If $b>1$ and $d>b$, then  $\lim_{t\to \infty}\phi=\infty$ and $\lim_{t\to \infty} x=0$, and\\
2) If $b<-1$ and $d<b$, then $\lim_{t\to \infty}\phi=0$ and $\lim_{t\to \infty} x=0$.\\
This is consistent with the corresponding limit (\ref{eq40}) for the $\Gamma$-coordinate of the point {\bf A6}.
The de Sitter solution for the point  {\bf A6} is the same obtained for the point {\bf A5}, given by the Eqs. (\ref{eq41}) and (\ref{eq43}), with the same limits for the scalar field and the $x$ and $\Gamma$ coordinates.\\
%%%%%%%%%%%%%%%%%%%%%%%%%%%%%%%%%%%%%%%%%%%%%%%%%%%%%%%%%%
\noindent Let' s turn to the case with exponential couplings and analyze the behavior at the coordinates of the critical points. Evaluating $\epsilon$ given in (\ref{eq26}) at the fixed point {\bf B5} one finds from the last equation in (\ref{eq7})
\be\label{eq49}
\dot{H}=-\frac{(b-c)(2b-c)}{b(b+c)}
\ee
leading to the solution for the scale factor
\be\label{eq50}
a=a_0(t-t_0)^{\gamma},\;\;\; \gamma=\frac{b(b+c)}{(b-c)(2b-c)}
\ee
In the phantom case (negative power) one can write $a=a_0(t_c-t)^{-|\gamma|}$. From the dynamical variables $f$ and $\Gamma$ (see (\ref{eq7}) and (\ref{eq16})) evaluated at the fixed point {\bf B5} one finds
\be\label{eq51}
\frac{f}{\Gamma}\Big |_{B5}=\frac{4b-2c}{(b+c)b}=\frac{\dot{\phi}}{H\phi}.
\ee
and after integration
\be\label{eq52}
\phi=\phi_0(t-t_0)^{\frac{2}{b-c}}.
\ee
These expressions allow us to analyze the behavior of the coordinate $x$ at $t\rightarrow \infty$:
\be\label{eq53}
x\propto \frac{(t)^{\frac{4}{b-c}}}{\kappa^{-2}-\xi e^{b\phi_0 (t)^{\frac{2}{b-c}}}},
\ee
where we have used $h(\phi)=\xi e^{b\phi}$, with the following limits:\\
1) If $b<c$, then $\lim_{t\to \infty}\phi=0$ and $\lim_{t\to \infty} x=0$, and\\
2) If $b>c$, then  $\lim_{t\to \infty}\phi=\infty$ and $\lim_{t\to \infty} x=0$.\\
Analyzing the $\Gamma$-coordinate we find two ways of getting consistent limits for the coordinates of the critical point: \\
\be\label{eq53a}
\Gamma\Big |_{B5}=\lim_{t\to \infty}\left(\frac{-\xi b e^{b\phi}}{\kappa^{-2}-\xi e^{b\phi}}\right)
\ee
1) If $b>c$, then $\lim_{t\to \infty}\Gamma=b$, which is compatible with the restrictions discussed at the end of the point {\bf B5} for the existence of stable quintessence or phantom solutions.\\
2) In the case $b<c$  ($\lim_{t\to \infty}\phi=0$), the limit 
$\Gamma\rightarrow b$ is valid in the approximation of the strong coupling limit when $\xi>>\kappa^{-2}$, and therefore the stable quintessence and phantom solutions can be considered in this limit for $b<c$. In this limit the effective Newtonian coupling becomes constant.\\
The de Sitter solution for the point {\bf B5} is obtained for $b=c$, and the values of the corresponding coordinates at this point lead to the solutions as follows. The Hubble parameter is constant and the scale factor is an exponential function as given by the Eq. (\ref{eq41}). The relation $f/\Gamma$ at this point gives (see (\ref{eq42}))
\be\label{eq54}
\phi=\phi_0 e^{\frac{H_0}{b}(t-t_0)}
\ee
which imply for the coordinate $x$ 
\be\label{eq55}
x\propto \frac{e^{\frac{2H_0}{b}(t-t_0)}}{\kappa^{-2}-\xi e^{b\phi_0 e^{\frac{H_0}{b}(t-t_0)}}}
\ee
and at $t\to \infty$ we can deduce \\
1) If $b>0$, then $\lim_{t\to \infty}\phi=\infty$ and $\lim_{t\to \infty} x=0$, and\\
2) If $b<0$, then $\lim_{t\to \infty}\phi=0$ and $\lim_{t\to \infty} x=0$.\\
After replacing the scalar field (\ref{eq54}) in the expression (\ref{eq53a}) for the $\Gamma$-coordinate, one finds, for $b>0$, $\lim_{t\to \infty}\Gamma=b$, and the effective Newtonian coupling vanishes at this limit. In the case $b<0$, at ${t\to \infty}$ we can consider the approximation of the strong coupling limit where $\xi>>\kappa^{-2}$, which leads to $\Gamma\to b$, and the effective Newtonian coupling becomes constant at this limit.\\
%%%%%%%%%%%%%%%%%%%%%%%%%%%%%%%%%%%%%%%%%%%%%%%
Proceeding in the same way with the fixed point {\bf B6} we find
\be\label{eq56}
a=a_0(t-t_0)^{\frac{b+d}{d-b}}, \;\;\; \phi=\phi_0(t-t_0)^{\frac{2}{d-b}}
\ee
Analyzing the $x$-coordinate at large times it's found
\be\label{eq57}
x\propto \frac{t^{\frac{4}{d-b}}}{\kappa^{-2}-\xi e^{b\phi_0 t^{\frac{2}{d-b}}}}
\ee
with the following limits:\\
1) If $b<d$, then  $\lim_{t\to \infty}\phi=\infty$ and $\lim_{t\to \infty} x=0$, and\\
2) If $b>d$, then $\lim_{t\to \infty}\phi=0$ and $\lim_{t\to \infty} x=0$.\\
This leaves for the $\Gamma$-coordinate, in the case $b<d$, $\lim_{t\to \infty}\Gamma=b$ with vanishing effective Newtonian coupling, and in the case $b>d$  ($\lim_{t\to \infty}\phi=0$), in the approximation of the strong coupling limit we find $\Gamma\rightarrow b$, and the effective Newtonian coupling tends to constant value. The scalar field for the de Sitter solution in {\bf B6} is the same as the one obtained for the point {\bf B5} (\ref{eq54}) and the coordinates of the fixed point have the same asymptotic behavior with the same consequences for the effective Newtonian coupling. 
%%%%%%%%%%%%%%%%%%%%%%%%%%%%%%%%%%%%%%%%%%%%%%%%%%
%%%%%%%%%%%%%%%%%%%%%%%%%%%%%%%%%%%%%%%%%%%%%%%%
\section{Discussion}
The scalar-tensor models represent a good source for modeling the dark energy and indeed, the explanation of the accelerated expansion of the Universe. In this regard, it is important to ask about the relevancy of the scalar-tensor couplings, predicted by fundamental theories, at current low-curvature universe.\\
%The couplings of scalar field to curvature appear in a variety of contexts 
%ranging from the quantization of matter fields on curved background to multidimensional theories like string or M-theory.
In the present work we studied some aspects of the late-time cosmological dynamics for the scalar-tensor model with non-minimal and Gauss-Bonnet couplings (see Eqs. (\ref{eq1}) and (\ref{eq2})). We considered the autonomous system and analyzed the critical points for two types of couplings and potential: for power-law couplings $h(\phi)\propto \phi^{b+1}$, $\eta(\phi)\propto \phi^{d+1}$ and potential $V(\phi)\propto \phi^c$ and for exponential couplings and potential $h(\phi)\propto e^{b\phi}$, $\eta(\phi)=\propto e^{d\phi}$ and $V(\phi)=\propto e^{c\phi}$. 
The presence of the GB coupling gives additional solutions with respect to the model of scalar field with non-minimal coupling that has been already considered in \cite{msami}, for power-law functions of the scalar field for the non-minimal coupling and potential.
In the case of power-law functions of the scalar field for the couplings and potential, we have described nine critical points, two of which we highlight here, the points {\bf A5} and {\bf A6}, since they contain stable quintessence and phantom solutions besides the stable de Sitter solutions. The critical point {\bf A5} becomes a de Sitter solution under the restriction $c=b+1$, and the stability depends on the relation between $b\ne 1$ and $d$ as discussed in the point {\bf A5}. Particularly the case $b=1$, which gives the standard non-minimal coupling $\xi \phi^2$, leads to de Sitter solution with marginal stability since one of the eigenvalues is zero (the others are negative), and the Higgs-like potential ($V\propto \phi^4$) leads to stable de Sitter expansion. This point (dominated by the scalar field) can also describe stable solutions with equation of state for the dark energy in the region around $w_{DE}=w_{eff}=-1$, with values above or below $-1$, corresponding to quintessence and phantom behavior respectively. It is worth noting that the limit of the $\gamma$-coordinate ($\lim_{t\to \infty}\Gamma=b+1$) is reached only at the large non-minimal coupling limit ($\phi^{b+1}>>1$). This limit is achieved for $\beta>0$, $b>1$ (in this case  $\lim_{t\to \infty}\phi=\infty$), or $\beta<0$, $b<-1$ (in this case $\lim_{t\to \infty}\phi=0$), and for both cases the effective Newtonian constant, defined as $F(\phi)^{-1}$, vanishes at the critical point. The combined effect of the non-minimal and GB couplings is reflected in the point {\bf A6} where the effective EoS depends on the two parameters $b$ and $d$, tough the stability involves the $c$-parameter of the potential. For this point the de Sitter solution is reached when $d=b$, which is stable whenever $c<b$ or saddle if $c>b$. Particularly the potentials: $V=const$, $V\propto \phi^2$ and $V\propto \phi^4$ give stable de Sitter solutions. The case $b=1$ leads to marginally stable de Sitter solution with one zero-eigenvalue as in the point {\bf A5}. This point also describes stable quintessence and phantom solutions as discussed in {\bf A6}. In this point the $\lim_{t\to \infty}\Gamma=b+1$, is reached also at $\phi\to 0$ and $\phi\to \infty$, but at both limits the effective Newtonian coupling vanishes. 
The point {\bf A7} contains an interesting scaling solution for the radiation dominated universe, with the scalar field being subdominant. With $w_m=1/3$ this point is saddle and can be considered as a transient phase of radiation dominated universe.\\
The exponential functions of the scalar field for the couplings and the potential, which are typical of string-inspired gravity models, give rise to new critical points that contain stable quintessence and phantom solutions, including also de Sitter solutions. The critical point {\bf B5} contains a de Sitter solution for $b=c$, which is an attractor node for $b>d$ and saddle for $b<d$. The consistency with the coordinates of the fixed point in the case $b>0$ leads to the vanishing of the effective Newtonian coupling, while in the case $b<0$ this effective coupling tends to a constant value at $t\to \infty$. Similar behavior happens for the quintessence and phantom solutions, where the effective Newtonian coupling vanishes for $b>c$ at $t\to \infty$, and for $b<c$ becomes constant. \\
The point {\bf B6} reflects the combined effect of the non-minimal and GB couplings and leads to stable de sitter, quintessence or phantom scenarios. For $b=d$ the system reaches a de Sitter fixed point, which is stable in the case $b>c$ and saddle if $b<c$. 
The coordinates of the de Sitter solution have exactly the same asymptotic behavior with the same consequences for the effective Newtonian coupling that the point {\bf B5}. Analyzing the asymptotic behavior of the effective Newtonian coupling, for quintessence and phantom scenarios, we found that it vanishes for $b<d$ and becomes constant for $b>d$. 
Additionally, the points {\bf B5} and {\bf B6} also give scaling solutions with dominance of the scalar field, i.e. $\Omega_{\phi}$ is not subdominant, contrary to what we would expect in early time radiation or matter dominated universe. The point {\bf B7} describes the same scaling solution for the radiation dominated universe that the point {\bf A7}.\\
An important difference between the power-law and exponential models is that in the last case the existence of quintessence, phantom or de Sitter solutions, allows an asymptotic behavior where the effective Newtonian coupling becomes constant. Another advantage of the exponential functions is that, given the fact that the parameters $b$, $c$ and $d$ take real values, we can adjust the EoS of the dark energy to asymptotic values as close to $-1$ as required. For the power-law functions these parameters were restricted to take integer values. Additionally, in all the above solutions the phantom scenario could be realized without introducing ghost degrees of freedom, which is quite attractive for a viable model of dark energy.
In the present analysis we have shown that the effect of the non-minimal and GB couplings lead to very interesting cosmological scenarios that can account for different accelerating regimes of the universe.
%Finally, also in this case, the high dimensionality of the phase space prevents an effective
%graphical description of the phase space and we will not include them here..............
%n all cases we have found that the critical points cover a variety of DE scenarios with EoS in the range of the measurements %made by the different collaborations.\\
%%%%%%%%%%%%%%%%%%%%%%%%%%%%%%%%%%%%%%%%%%%%%%%%%%%
\section*{Acknowledgments}
This work was supported by Universidad del Valle under project CI 71074, DFJ acknowledges support from COLCIENCIAS, Colombia.
%\nocite{*}

\begin{thebibliography}{99}
\bibitem{riess} A.G. Riess, \textit {et al.}, Astron. J. {\bf 116}, 1009 (1998); astron. J. {\bf 117},
707 (1999). 
\bibitem{perlmutter} S.Perlmutter \textit{et al.}, Nature \textbf{391}, 51 (1998)
\bibitem{kowalski} M. Kowalski, \textit{et. al.}, Astrophys. Journal, \textbf{686}, p.749 (2008), arXiv:0804.4142
\bibitem{hicken} M. Hicken et al., Astrophys. J. 700, 1097 (2009) [arXiv:0901.4804 [astro-ph.CO].
\bibitem{komatsu} E. Komatsu et al. [WMAP Collaboration], Astrophys. J. Suppl. 180, 330 (2009)
arXiv:0803.0547 [astro-ph].
%\bibitem{wmap} WMAP Collaboration (Komatsu E. \textit{et al.}), Astrophys. J. Suppl., 180, (2009) 330.
\bibitem{percival} Percival W. J. \textit{et al.}, Mon. Not. R. Astron. Soc., {\bf 401} (2010), 2148.
\bibitem{planck1} Planck Collaboration (Ade P. A. R. \textit{et al.}), arXiv:1303.5062.
\bibitem{planck2} Planck Collaboration (Ade P. A. R. \textit{et al.}), arXiv:1303.5076.
\bibitem{starovinsky1} V. Sahni, A. Starobinsky, Int. J. Mod. Phys. D 9, 373 (2000); arXiv:astro-ph/9904398
\bibitem{peebles} P. J. E. Peebles and B. Ratra, Rev. Mod. Phys. 75, 559 (2003) [arXiv:astroph/0207347]
\bibitem{padmanabhan} T. Padmanabhan, Phys. Rept. 380, 235 (2003) [arXiv:hep-th/0212290]
\bibitem{chiba} T. Chiba, Phys.Rev. D60, 083508 (1999); gr-qc/9903094.
\bibitem{uzan} J.-P. Uzan, Phys. Rev. D 59, 123510 (1999); gr-qc/9903004.
\bibitem{lamendola} L. Amendola, Phys.Rev. D60, 043501, (1999); astro-ph/9904120.
\bibitem{boisseau} B. Boisseau, G. Esposito-Farese, D. Polarski, A. A. Starobinsky, Phys.Rev.Lett. 85, 2236 (2000); gr-qc/0001066.
\bibitem{polarski} G. Esposito-Farese, D. Polarski, Phys. Rev. D 63, 063504 (2001);  gr-qc/0009034.
\bibitem{sergei12} S. Nojiri, S. D. Odintsov and M. Sasaki, Phys. Rev. \textbf{D71}, 123509 (2005); hep-th/0504052.
\bibitem{koivisto2} T. Koivisto, D. F. Mota, Phys. Rev. \textbf{D75}, 023518 (2007); hep-th/0609155
\bibitem{carloni1} S. Carloni, J. A. Leach, S. Capozziello, P. K. S. Dunsby, Class. Quant. Grav. \textbf{25}, 035008 (2008); gr-qc/0701009
\bibitem{sushkov} S.V. Sushkov, Phys. Rev. D80, 103505 (2009); arXiv:0910.0980
\bibitem{saridakis} E. N. Saridakis, J. M. Weller, Phys. Rev. \textbf{D81}, 123523 (2010); arXiv:0912.5304 [hep-th]
\bibitem{granda1} L. N. Granda, JCAP \textbf{07}, 006 (2010); arXiv:0911.3702 [hep-th]
\bibitem{saridakis1} E.N.Saridakis, S.V.Sushkov, Phys. Rev. D81, 083510 (2010); arXiv:1002.3478
\bibitem{granda2} L. N. Granda and W. Cardona, JCAP \textbf{07}, 021 (2010); arXiv:1005.2716 [hep-th]
\bibitem{granda3} L. N. Granda, Int. J. Theor. Phys. 51 (2012) 2813; arXiv:1109.1371 [gr-qc].
\bibitem{sushkov1} M. A. Skugoreva, S. V. Sushkov, A. V. Toporensky, Phys. Rev. \textbf{D88}, 083539 (2013); arXiv:1306.5090 [gr-qc] 
                         %%%%%%%%%%%%%%%%%%%%%%%%%%%%%%%%
\bibitem{capozziello1} S. Capozziello, Int. J. Mod. Phys. D 11, 483 (2002).
\bibitem{capozziello2} S. Capozziello, V. F. Cardone, S. Carloni and A. Troisi, Int. J. Mod. Phys. D, \textbf{12}, 1969 (2003).
\bibitem{sergei1} S. Nojiri and S. D. Odintsov, Phys. Rev. D 68, 123512 (2003).
\bibitem{carroll} S. M. Carroll, V. Duvvuri, M. Trodden and M. S. Turner, Phys. Rev. \textbf{D70}, 043528 (2004).
\bibitem{sergei2} S. Nojiri and S. D. Odintsov, Phys. Rept. \textbf{505}, 59 (2011); arXiv:1011.0544 [gr-qc]
\bibitem{sotiriou} T. P. Sotiriou, V. Faraoni, Rev. Mod. Phys. \textbf{82}, 451 (2010); arXiv:0805.1726 [gr-qc]
\bibitem{tsujikawa} S. Tsujikawa, Lect. Notes Phys. 800, 99 (2010);  arXiv:1101.0191 [gr-qc] 
\bibitem{copeland} E. J. Copeland, M. Sami and S. Tsujikawa, Int. J. Mod. Phys. D \textbf{15}
1753-1936 (2006), arXiv:hep-th/0603057
\bibitem{sahnii} V. Sahni, 	Lect. Notes Phys. \textbf{653}, 141-180 (2004), arXiv:astro-ph/0403324v3
\bibitem{sergeiod} K. Bamba, S. Capozziello, S. Nojiri, S. D. Odintsov, Astrophys.  and Space Sci. {\bf 342}, 155  (2012); arXiv:1205.3421 [gr-qc] 
\bibitem{nojiriod} S. Nojiri and S. D. Odintsov, Phys. Rept. 505, 59 (2011), arXiv:1011.0544 [gr-qc].
                          %%%%%%%%%%%%%%%%%%%%%%%%%%%%%%%%
 \bibitem{metsaev} R. Metsaev, A. Tseytlin, Nucl. Phys. B {\bf 293}, 385 (1987)                         
 \bibitem{meissner}  K. A. Meissner, Phys. Lett. B392, 298 (1997), arXiv:hep-th/9610131 [hep-th].                       
 \bibitem{ford} L.H. Ford, Phys. Rev. D35, 2955 (1987)
\bibitem{birrel} N.D. Birrell and P.C.W. Davis, {\it Quantum fields
in curved spacetime} (Cambridge University Press) (1982) 
\bibitem{lamendola1} L. Amendola, C. Charmousis, S. C.  Davis, JCAP  {\bf 0612}, 020 (2006);
arXiv:hep-th/0506137
\bibitem{perivolaropoulos} L. Perivolaropoulos, JCAP 0510, 001 (2005), arXiv:astro-ph/0504582 [astro-ph].
\bibitem{fujii} Y. Fujii and K. Maeda, The scalar-tensor theory of gravitation (Cambridge University Press, 2007).
%\bibitem{uzan} J.-P. Uzan, Phys. Rev. D 59, 123510 (1999); gr-qc/9903004.
%\bibitem{lamendola} L. Amendola, Phys.Rev. D60, 043501 (1999); astro-ph/9904120.
\bibitem{perrotta} F. Perrotta, C. Baccigalupi and S. Matarrese, Phys. Rev. D 61, 023507 (2000); astro-ph/9906066.
\bibitem{riazuelo} A. Riazuelo and J.-P. Uzan, Phys. Rev. D62, 083506 (2000); astro-ph/0004156.
\bibitem{perrotta1} C. Baccigalupi, S. Matarrese, F. Perrotta, Phys.Rev. D62, 123510 (2000);  astro-ph/0005543 
\bibitem{capozziello3} S. Capozziello, S. Nesseris, L. Perivolaropoulos, JCAP 0712 (2007) 009;  arXiv:0705.3586 [astro-ph]
\bibitem{tchiba} T. Chiba, Phys. Rev. D 64, 103503 (2001); astro-ph/0106550.
\bibitem{vfaraoni} V. Faraoni, Int. J. Mod. Phys. D 11, 471 (2002);  astro-ph/0110067.
\bibitem{elizalde} E. Elizalde, S. Nojiri, S. D. Odintsov, Phys. Rev. D 70, 043539 (2004), hep-th/0405034
\bibitem{nojiri3} S. Nojiri, E. N. Saridakis, Astrophys. Space Sci. 347, 221 (2013);  arXiv:1301.2686 [hep-th] 
\bibitem{carvalho} F. C. Carvalho, A. Saa, Phys. Rev. D70, 087302  (2004); arXiv:astro-ph/0408013.
\bibitem{polarski1} R. Gannouji, D. Polarski, A. Ranquet, A. A. Starobinsky, JCAP 0609, 016 (2006); astro-ph/0606287 
\bibitem{vfaraoni1} V. Faraoni, Phys. Rev. D 70, 044037 (2004); gr-qc/0407021 
\bibitem{msami} M. Sami, M. Shahalam, M. Skugoreva, A. Toporensky, Phys. Rev. D 86,103532 (2012); arXiv:1207.6691 [hep-th]
\bibitem{sergei4} S. Nojiri, S D. Odintsov, M. Sasaki, Phys. Rev. D71, 123509 (2005); arXiv:hep-th/0504052.
\bibitem{tsujikawa1} S. Tsujikawa and M. Sami, JCAP  0701 (2007) 006; arXiv:hep-th/0608178.
\bibitem{leith} B. M. Leith and I. P. Neupane, J. Cosmol. Astropart. Phys. 0705 (2007) 019; arXiv:hep-th/0702002.
\bibitem{koivisto1} T. Koivisto and D. F. Mota, Phys. Lett. B 644 (2007) 104; arXiv:astro-ph/0606078.
%\bibitem{koivisto2} T. Koivisto and D. F. Mota, Phys. Rev. D75, 023518 (2007), arXiv:hep-th/0609155 [hep-th].
\bibitem{neupane2} I. P. Neupane, Class. Quantum Grav. 23 (2006) 7493; arXiv:hep-th/0602097.
\bibitem{nojiriod1} S. Nojiri and S. D. Odintsov, Phys. Lett. B 631 (2005) 1 , arXiv:hep-th/0508049.
\bibitem{granda5} L. N. Granda and E. Loaiza, Int. J. Mod. Phys. D2, 1250002 (2012), arXiv:1111.2454 [hep-th].
\bibitem{granda6} L. N. Granda, Int. J. Theor. Phys. 51, 2813 (2012); arXiv:1109.1371 [gr-qc].
\bibitem{granda7} L. N. Granda, D. F. Jimenez, Phys. Rev. D {\bf 90}, 123512 (2014); arXiv:1411.4203 [gr-qc].

%%%%%%%%%%%%%%%%%%%%%%%%%%%%%%%%%%%

\end{thebibliography}

\end{document}